\documentclass[fleqn]{2023SCGE}
\usepackage{amsmath}
\usepackage{threeparttable}
\usepackage{ulem}

\newcommand{\oiii}{[O\,{\sc iii}]}

\newcommand{\oi}{[O\,{\sc i}]}
\newcommand{\hb}{H$\beta$}
\newcommand{\ha}{H$\alpha$}
\newcommand{\nii}{[N\,{\sc ii}]}
\newcommand{\sii}{[S\,{\sc ii}]}
\newcommand{\greenpea}{{\it Green pea}}
\newcommand{\blue}{{\it Blueberry}}
\newcommand{\purple}{{\it Purple grape}}
\newcommand{\MBH}{$M_{\rm BH}$}
\newcommand{\Mstar}{$M_{\rm *}$}
\newcommand{\Msun}{$\rm M_{\odot}$}
\begin{document}

\ensubject{subject}

\ArticleType{Article}
\SpecialTopic{SPECIAL TOPIC: }
\Year{2024}
\Month{xxx}
\Vol{xx}
\No{x}
\DOI{??}
\ArtNo{000000}

\title{Intermediate-Mass Black Holes in Green Pea Galaxies (IMBH-GP) I: a Candidate Sample from LAMOST and SDSS}

\author[1,2]{Ruqiu Lin}{}%
\author[1,2]{Zhen-Ya Zheng}{{zhengzy@shao.ac.cn}}
\author[1]{Fang-Ting Yuan}{}
\author[3]{Jun-Xian Wang}{}%
\author[1]{Chunyan Jiang}{}
\author[3]{Ning Jiang}{}
\author[4,5]{\\Lingzhi Wang}{}
\author[6,7]{Linhua Jiang}{}
\author[1]{Xiang Ji}{}
\author[1,2]{Shuairu Zhu}{}
\author[1]{Xiaodan Fu}{}

\AuthorMark{Lin R. Q., Zheng Z.-Y., Yuan F.-T.}

\AuthorCitation{Lin R. Q., Zheng Z.-Y., Yuan F.-T., Wang J.-X., Jiang C. Y., Jiang N., Wang L. Z., Jiang L. H., Ji X., Zhu S. R., and Fu X. D.}

\address[1]{Key Laboratory for Research in Galaxies and Cosmology, Shanghai Astronomical Observatory, Chinese Academy of Sciences, Shanghai {\rm 200030}, China}
\address[2]{School of Astronomy and Space Sciences, University of Chinese Academy of Sciences, Beijing {\rm 100049}, China}
\address[3]{CAS Key Laboratory for Research in Galaxies and Cosmology, Department of Astronomy, University of Science and Technology of China, Hefei {\rm 230026}, China}
\address[4]{Chinese Academy of Sciences South America Center for Astronomy (CASSACA), National Astronomical Observatories, Chinese Academy of Sciences, Beijing {\rm 100101}, China}
\address[5]{CAS Key Laboratory of Optical Astronomy, National Astronomical Observatories, Chinese Academy of Sciences, Beijing {\rm 100101}, China}
\address[6]{Kavli Institute for Astronomy and Astrophysics, Peking University, Beijing 100871, China}
\address[7]{Department of Astronomy, School of Physics, Peking University, Beijing 100871, China}


\abstract{
The scaling relation of central massive black holes (MBHs) and their host galaxies is well-studied for supermassive BHs (SMBHs, \MBH$\ \ge 10^6\,$\Msun). 
However, this relation has large uncertainties in the mass range of the intermediate-mass BHs (IMBHs, \MBH$\ \sim10^3-10^{6}\,$\Msun).
Since \greenpea\ (GP) galaxies are luminous compact dwarf galaxies, which may be likely to host less massive SMBHs or even IMBHs, we systematically search for MBHs in a large sample of 2190 GP galaxies at $z < 0.4$, selected from LAMOST and SDSS spectroscopic surveys. Here, we report a newly discovered sample of 59 MBH candidates with broad \ha\ lines.
This sample has a median stellar mass of $10^{8.83\pm0.11}\,$\Msun\ and hosts MBHs with single-epoch virial masses ranging from \MBH$\ \sim 10^{4.7}$ to $10^{8.5}\, $\Msun\ (median $10^{5.85\pm0.64}\,$\Msun). Among the 59 MBH candidates, 36 have black hole masses \MBH\ $\le 10^{6}\,$\Msun\ (IMBH candidates), one of which even has \MBH$\ \lesssim 10^{5}\,$\Msun.
We find that the \MBH$-$\Mstar\ relation of our MBH sample is consistent with the \MBH$-M_{\rm bulge}$ relation for SMBHs, while is above the \MBH$-$\Mstar\ relation for MBHs in dwarf galaxies in the same mass range.
Furthermore, we show that 25 MBH candidates, including 4 IMBH candidates, have additional evidence of black hole activities, assessed through various methods such as the broad-line width, BPT diagram, mid-infrared color, X-ray luminosity, and radio emission.
Our studies show that it is very promising to find IMBHs in GP galaxies, and the BH sample so obtained enables us to probe the connection between the MBHs and compact dwarf galaxies in the low-redshift Universe. 
}
\keywords{SMBHs, IMBHs, Green Pea Galaxies, Compact Dwarf Galaxies}

\PACS{95.80.+p, 98.54.-h, 98.54.Ep, 98.62.Ai, 98.62.Js}

\maketitle


\begin{multicols}{2}
\section{Introduction} 
\label{sec:intro}
The co-evolution of galaxies and central massive black holes (MBHs) has been an important topic in galaxy research, with a tight correlation observed between black hole (BH) masses and stellar masses of host galaxies. This correlation is initially established in supermassive black holes (SMBHs) and their host galaxies \cite{Kormendy2013}. However, a longstanding challenge exists in the scarcity of intermediate-mass black holes (IMBHs, $10^{3} <\ $\MBH$\ < 10^{6}\,$\Msun \footnote{We note that ref. \cite{Greene2020} defined the mass range of IMBHs as $10^{3} <\ $\MBH$\ < 10^{5}\,$\Msun. But they also include the MBHs with \MBH$\ \sim 10^{5} - 10^{6}\,$\Msun\ in their discussion. For consistency, the IMBHs in this paper are referred to as MBHs with $10^{3} <\ $\MBH$\ < 10^{6}\,$\Msun.} \cite{Greene2004, Greene2007b, Mezcua2017, Shin2022}). 
Besides, inspired by observational evidence of high-redshift SMBHs, theoretical predictions suggest that IMBHs can be seeds of early-Universe SMBHs. Hence, in the local universe, dwarf galaxies formed in the early Universe with minimal environmental impact can host leftover seed BHs \cite{Mezcua2017}. 

Subsequent studies have explored IMBH searches in nearby dwarf galaxies \cite{Ho1997, Greene2004, Greene2007, Dong2012, Reines2013}, determining black hole virial masses (\MBH$\ \propto R\Delta V^2/G$) by extracting weak broad \ha\ signals from spectra. This conversion relation, derived from the broad-line region size-luminosity relation, provides a relationship between broad \ha\ lines and black hole mass \cite{Greene2005} based on empirical relations for different Balmer emission lines (\ha\ and \hb) luminosities. 
In the past decades, there have been substantial efforts to search for MBHs in dwarf galaxies based on the single-epoch broad \ha\ technique \cite{Greene2007, Dong2012, Reines2013, Reines2015, Liu2018, Salehirad2022}. About 1000 MBH candidates have been found in dwarf galaxies so far.
However, few samples now have black hole masses less than $10^5\,$\Msun. A considerable uncertainty in the \MBH$-$\Mstar\ relation at the low-mass end of MBHs still exists. 

\greenpea\ (GP) galaxies are compact, dwarf galaxies with intense \oiii\ $\lambda$5007 emission lines in the local Universe \cite{Cardamone2009, Izotov2011, Yang2017, Brunker2020}.
They are regarded as analogs of those strong emission line galaxies (e.g., Ly$\alpha$ emitters and Lyman continuum galaxies) in early Universe \cite{Izotov2021A}, which may play an important role in ionizing the neutral hydrogen environment in the epoch of reionization \cite{Ouchi2020, Izotov2016A, Izotov2016B, Izotov2021B}. Recently, JWST has conﬁrmed the existence of GPs at very high redshift \cite{Schaerer2022, Taylor2022, Trump2023, Curti2023, Rhoads2023}. In the meanwhile, JWST/NIRSpec and NIRcam deep spectroscopic observations have confirmed dozens of active galactic nuclei (AGNs) in dwarf galaxies at $4<z<9$, which have the single-epoch virial BH mass in the range of $10^5 - 10^9\, $\Msun\ \cite{Maiolino2023, Harikane2023, Matthee2023, Greene2023, Kocevski2023, Kokorev2023, Larson2023, Furtak2023, Yue2023}. Those AGNs appear as ``red dots" with a reddened continuum in the optical band and compact morphology. The connection between the compact morphology and BH activity remains unclear.  
Therefore, GP galaxies can help us understand not only the escape mechanisms of both Ly$\alpha$ and Lyman continuum photons but also the properties of MBHs in this special sample as analogs of high-redshift compact galaxies.

Luminous, compact dwarf galaxies like GP galaxies are likely to host IMBHs instead of SMBHs \cite{Izotov2011}. 
However, in previous studies, black hole samples in GP galaxies have not received much attention and are typically excluded in various analyses. Some studies have pointed out that low-metallicity AGNs in GP galaxies can be overlooked by BPT (Baldwin-Phillips-Terlevich, \cite{Baldwin1981}) classification \cite{Izotov2011}. WISE infrared color and variability analyses suggest a certain proportion of missed AGNs in GP galaxies \cite{Harish2023}. In our earlier work \cite{Lin2023}, we report research on five AGNs with double-peaked narrow emission lines in GP galaxies. Their double-peaked emission line profiles may originate from a dual AGN merger event, suggesting that the extreme \oiii\ emission lines in GP galaxies may be associated with binary BH mergers.
However, there is currently a lack of systematic research on identifying MBHs in GP galaxies.

This is the first paper in a series dedicated to the study of the properties of MBHs in GP galaxies at $z<0.4$ from LAMOST and SDSS.
In this paper, we report a new sample of MBH candidates selected from a large sample of GP galaxies from LAMOST and SDSS surveys. We model the emission-line profiles, obtain the broad \ha\ component, and finally estimate the single-epoch virial BH masses of GPs with a broad \ha\ component. We then construct the mass distribution of IMBHs in the strong emission-line galaxies. In the second paper (Lin et al. in prep.) of this series, we reveal the connection between MBHs in GP galaxies at $z<0.4$ from LAMOST and SDSS and broad-line AGNs (including the Little Red Dots, LRDs) at $z>4$ from JWST.

Our paper is organized as follows.
We describe the data, the spectral analysis, and the sample selection in Section \ref{sec:data_sample} and present measurements of galaxy masses and BH masses in Section \ref{sec:mass_mea}. We describe our AGN identification in the candidate MBH sample in Section \ref{sec:AGN_id}. In Section \ref{sec:discussion}, we discuss the \MBH$-$\Mstar\ relation of GP galaxies and the contamination from supernovae (SNe) and tidal disruption events (TDEs) and summarize our conclusions in Section \ref{sec:conclu}. 

Throughout this paper, we assume the cosmological parameters of $H_0$ = 70 $\rm km\, s^{-1}\, Mpc^{-1}$, $\Omega_m$ = 0.3, and $\Omega_{\Lambda}$ = 0.7.

\section{Data and Sample} \label{sec:data_sample}
\label{sec:data}

\subsection{Parent Sample: \greenpea\ galaxies}
\label{sec:parentSample}
Our parent GP sample is composed of 2190 compact \oiii\ emission-line galaxies at $z <$ 0.4 obtained from the Large Sky Area Multi-Object Fiber Spectroscopic Telescope (\href{http://www.lamost.org/}{LAMOST}) DR9 and Sloan Digital Sky Survey (\href{https://www.sdss4.org/}{SDSS}) DR18. This sample contains not only GP galaxies at 0.12 $< z <$ 0.4 but also \purple\ and \blue\ galaxies at $z < 0.12$, which we refer to as GP galaxies as well.

The LAMOST sample contains 1547 GP galaxies, which is reported in ref. \cite{Liu2022}. This sample is mainly the input catalog of a PI project (PI: Junxian Wang) of the LAMOST extra-galactic survey add-on program, which aims to construct a non-biased and largest sample of GP galaxies to date. Briefly, this sample is selected with SDSS DR12 {\it ugriz} colors and compact sizes and covers the spectroscopic redshift from {\it z} = 0 to {\it z} = 0.72. We only retain sources with redshifts $z < 0.4$ to ensure that the spectra cover the Ha line. We also drop the spectra with no redshift measured from the LAMOST data reduction pipeline and 1466 galaxies are left for the following analyses.

The SDSS sample is comprised of 724 GP galaxies, which is confirmed with SDSS spectra. We first select the GP candidates based on the SDSS $grz$ color-color diagram. 
We select the GP candidates at 0.12 $\, < z <\, $ 0.4 following the criteria of ref. \cite{Cardamone2009}. For the candidates at $z <$ 0.12, the color criteria are 
$u-g \ge 0.4\, \&\, g-r \le -0.3$.
To ensure the compactness and brightness of this sample, we then restrict the GP candidates to those with {\tt petroRad\_r} $\le$ 5$\arcsec$ and 18 $\,< r <\,$ 20.5 mag.
We require {\tt mode = 1} to ensure the object is listed in the primary catalog.  

We also exclude the possible contaminants which are classified as {\it STAR} ({\tt TYPE} keyword) or labeled as {\it OBJECT1\_DEBLEND\_TOO\_MANY\_PEAKS}, {\it OBJECT1\_CR}, and {\it OBJECT1\_DEBLENDED\_AS\_PSF} in SDSS {\tt photoObjAll} catalog.

The LAMOST and SDSS fiber spectroscopes have comparable spectral resolving power of $R\sim1800$ and wavelength coverage of $\sim 3700-9000\,\AA$ (see ref. \cite{Du2016} for LAMOST and ref. \cite{York2000} for SDSS), corresponding to an instrumental dispersion of $\rm \sim69\, km\,s^{-1}$. The fiber diameters of LAMOST and SDSS are 3$\farcs$3 and 3\arcsec, respectively. 
Since the fluxes of LAMOST spectra are not physical, we follow ref. \cite{Wang2018} and ref. \cite{Liu2022} to re-calibrate the fluxes by comparing the magnitude of SDSS {\it gri} photometry and the re-calibrated fluxes are used throughout this paper.

\subsection{Emission-line Measurement}\label{sec:line_model}
We aim to extract the broad \ha\ component from optical spectra and then determine the virial BH masses of GP galaxies. We first subtract the stellar continuum and absorption lines to obtain the pure emission-line spectra. 
Then we fit \oiii\ lines with a single narrow component or a narrow component plus a wide component, where the narrow component represents the star formation and the possible narrow lines produced by AGNs, and the wide component indicates probable outflows \cite{Woo2016}. 
Finally, we check the existence of broad \ha\ components in the \ha\ line after subtracting the narrow and possible wide components fitted from the \sii\ or \oiii\ profile. The details are described below.

Due to the impact of stellar absorption on the measurement of broad components in Balmer lines, we first eliminate the stellar continuum from the spectrum. The continuum is fitted using {\tt pPXF} \cite{Cappellari2004, Cappellari2017, Cappellari2023} with spectral templates comprising the continuum and absorption-line spectra from stellar populations \cite{Vazdekis2016}, along with 21 gas emission lines. We then subtract the stellar emission to obtain the pure emission-line spectra.

Decomposing different components is crucial for searching for the broad \ha\ component. With pure emission-line spectra, we measure emission lines using the {\tt pyspeckit} \cite{Ginsburg2011, Ginsburg2022} package. 
The parameters of the narrow-line model are routinely determined by \sii\ $\lambda\lambda 6548,6583$ doublet (ref. \cite{Ho1997, Greene2004}, also see ref. \cite{Reines2013}).
We note that some GP galaxies show multiple components in \oiii\ lines, while \sii\ is not strong enough to decompose different spectra components. Therefore, we use \oiii\ $\lambda5007$\ as an alternative to \sii\ when the \oiii\ profile has more than one component.

Here we introduce the method of decomposing multiple components of \oiii.
We run the fitting twice, the first run using a single Gaussian for the narrow component (\oiii$\rm ^N$), and the second run adding a wide component (\oiii$\rm ^W$), hence referred to as the two-Gaussian model. 
We then determine which model to use, based on the reduced $\chi^2$, amplitude $Amp_{\rm [OIII]^W}$ of the wide component (if exists) of \oiii, root-mean-square flux $\sigma_{\rm rms}$ in the nearby line-free region, and the full-width at the half maximum ($FWHM$) of components of the fitting results in the above two models. We take into account the \oiii$\rm ^W$ component when the criteria below are met:
\begin{equation}
    \begin{split}
        \rm \chi^2_{1-Gaussian}/\chi^2_{2-Gaussian} > 1.2, \\
        Amp_{\rm [OIII]^W}/\sigma_{\rm rms} > 2, {\rm and}\\
        FWHM_{\rm [OIII]^W}>FWHM_{\rm [OIII]^N},
    \end{split}
\end{equation}
where subscripts $\rm _{1-Gaussian}$ and $\rm _{2-Gaussian}$ are for the single-Gaussian and two-Gaussian model, respectively. We adopt a laboratory ratio of 3 for \oiii\ $\lambda 5007$/\oiii\ $\lambda 4959$.
To derive the BPT diagram, we fit \hb\ together with \oiii. Because the \hb\ is much weaker than \oiii\ and \ha, we use a simple model with the same narrow component as \oiii\ and a wider component independent of \oiii. We also fit \oi$\lambda6300$ line with a single Gaussian profile to measure the emission-line flux.

We subsequently fit \ha-\nii\ complex and \sii\ lines in two cases: \oiii\ well fitted with the single-Gaussian model (\oiii$\rm ^N$) and \oiii\ well fitted with the two-Gaussian model (\oiii$\rm ^N$+\oiii$\rm ^W$). For the first case, we fit these lines as single-Gaussian profiles and fix the line widths to \sii\ doublet. 
For the second case, we fit each line with the same profile of \oiii, i.e., narrow (\ha$\rm ^N$ and \nii$\rm ^N$) + wide (\ha$\rm ^W$ and \nii$\rm ^W$) components. The width of each component and flux ratio and velocity difference between these two components are fixed according to \oiii\ but the flux values are independent. We adopt a laboratory ratio of 3 for \nii\ $\lambda 6583$/\nii\ $\lambda 6549$.
In search of the broad Ha component (\ha$\rm ^B$), next we fit the above emission lines again, by adding the \ha$\rm ^B$ component to each of the two cases.
The parameters of the \ha$\rm ^B$ component are independent. 
We caution that the components other than the broad component of \ha\ are determined by \oiii\ line profile, which may result in contamination of the broad \ha\ component when the wide component is NOT resolved in the \oiii\ fitting due to the limit of spectral resolution or/and sensitivity.
We then select the GP galaxies with potential broad \ha\ lines as our MBH candidates, described in Section \ref{sec:broad_criteria}.

\subsection{Broad \ha\ Candidates}
\label{sec:broad_criteria}
We select broad-line candidates in GP galaxies based on the reduced $\chi^2$, amplitude $Amp_{\rm H\alpha^W}$ of the broad \ha\ component (if exists), $\sigma_{\rm rms}$ in the nearby line-free region, and $FWHM$s of narrow and broad \ha\ components. Broad \ha\ candidates are those with:
\begin{equation}
    \begin{split}
        &1.\ \rm \chi^2_{w/o\ broad}/\chi^2_{w/\ broad} > 1.2 , \\
        &2.\ Amp_{\rm H\alpha^B}/\sigma_{\rm rms} > 2, {\rm and}\\
        &3.\ FWHM_{\rm H\alpha^B} > FWHM_{\rm H\alpha^N},
    \end{split}
\end{equation}
where the subscript $\rm _{w/\ broad}$ and $\rm _{w/o\ broad}$ are for the \ha\ model with and without broad component, respectively.

{\it Condition 1} requires the reduced $\chi^2$ to be improved by 20\% after adding a broad component, which is an empirical value and has proven effective in literature \cite{Hao2005, Dong2012, Reines2013}. {\it Condition} 2 is a significance cut of the broad \ha\ component (the same as Criterion 4 in ref.\cite{Dong2012}). Because we aim to identify weak broad Ha components, we  do not impose strict constraints on the line width of this component. Instead, we apply a relaxed restriction, as outlined in {\it Condition} 3. We also inspect all the fitting results and exclude five possible contaminants.

\begin{figure}[H]
    \centering
    \includegraphics[width=0.5\textwidth]{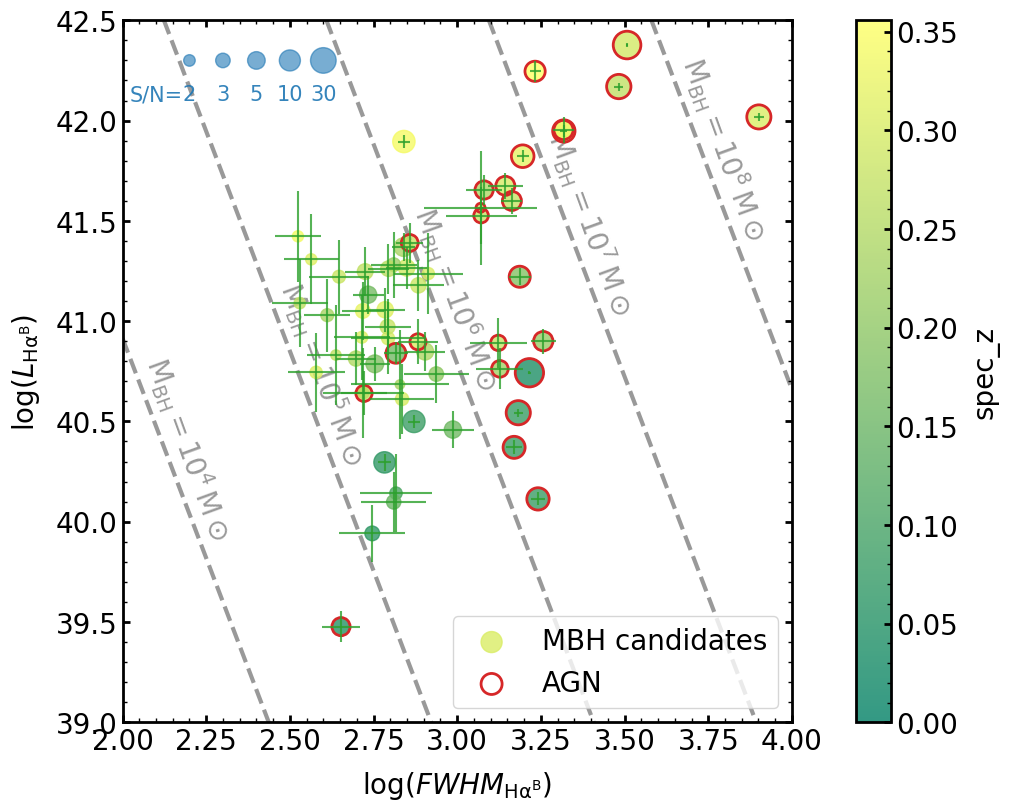}
    \caption{Luminosity as a function of FWHMs for the broad \ha\ components of MBH candidates in the sample of GP galaxies. The color bar shows the spectroscopic redshifts. The sizes of these dots represent the signal-to-noise ratio (S/N) of the broad \ha\ components. The error bars are the fitting error for the broad \ha\ components in the emission-line measurement. Furthermore, the identified AGNs are marked in red circles (see Section \ref{sec:AGN_id} in details).}
    \label{fig:LHa_FWHM}
\end{figure}

Finally, we obtain a sample of 59 MBH candidates, of which 37 and 22 are selected from SDSS and LAMOST, respectively. The lowest luminosity of broad \ha\ component we extract is $\rm \sim 10^{39.5}\, erg\, s^{-1}$ (see Figure~\ref{fig:LHa_FWHM}), which is about 30 times brighter than that of the nearest dwarf Seyfert galaxy NGC4395 hosting a known IMBH ($L_{\rm H\alpha^B} \sim 10^{38}\rm \, erg\, s^{-1}$ \cite{Filippenko1989}, a single-epoch virial BH mass of \MBH\ $\sim 10^{5}\,$\Msun\ \cite{LaFranca2015}, and a reverberation-mapping BH mass of \MBH\ $\sim 10^{4}\,$\Msun\ \cite{Woo2019}). We note that there is only one MBH candidate (J084029+470710) first reported as a BH candidate in other literature \cite{Izotov2007, Reines2013, Baldassare2016}.

We divide the MBH candidates into three subsets. Sample \textbf{A} (22) consists of candidates well-fitted with the two-Gaussian model of \oiii, suggesting the possible existence of outflows. Sample \textbf{B} (8) and sample \textbf{C} (29) are both well-fitted with the single-Gaussian model of \oiii, but when the narrow-line model changes to \sii, only the MBH candidates in sample \textbf{B} can extract the broad Ha components as well. Hence, we consider that sample \textbf{B} is more secure than sample \textbf{C}, as it does not depend on the narrow-line model.
 
\begin{figure*}[thbp]
    \centering
    \includegraphics[width=0.9\textwidth]{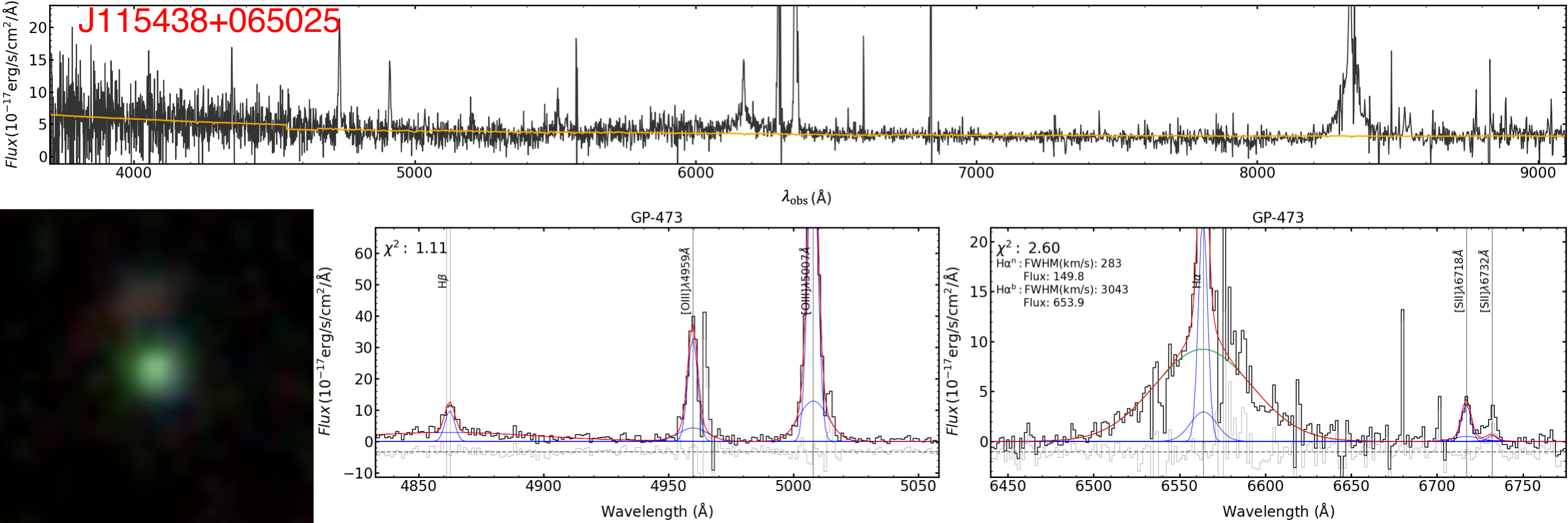}
    \includegraphics[width=0.9\textwidth]{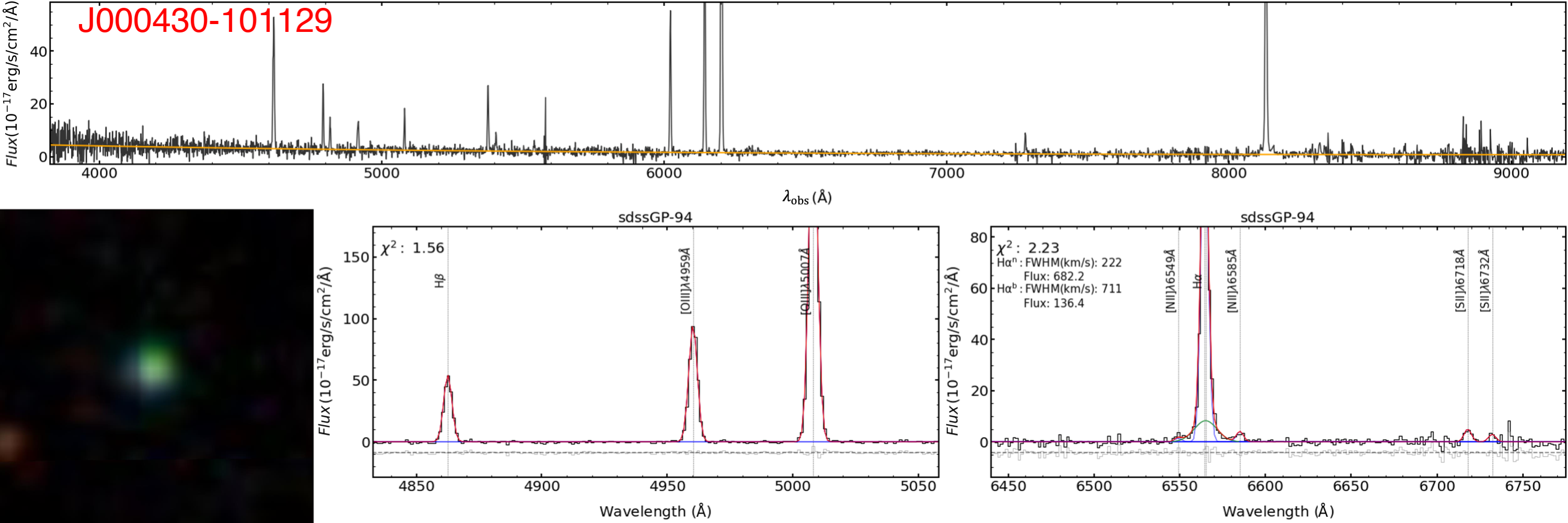}
    \includegraphics[width=0.9\textwidth]{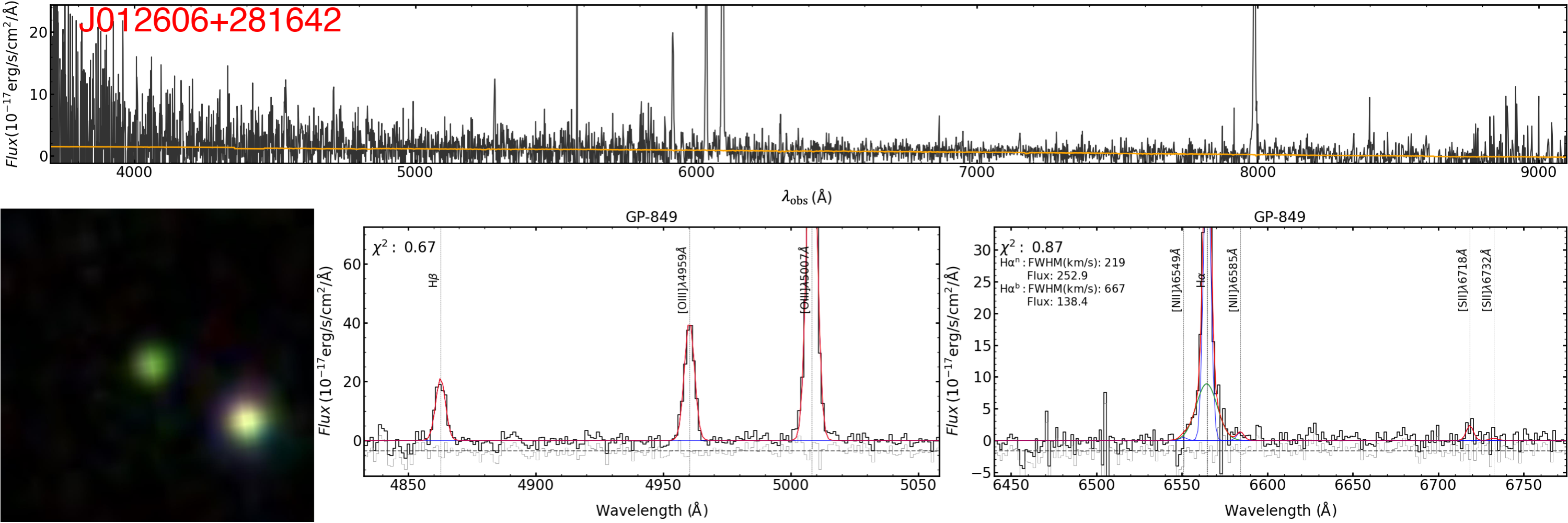}
    \caption{Examples of spectral analysis for the MBH candidates in the sample of GP galaxies. The rows from the top to bottom illustrate examples of sample \textbf{A}, sample \textbf{B}, and sample \textbf{C}. In each row, the top penal presents the fitting of stellar light. The left of the bottom penal presents the SDSS {\it gri} (blue, green, and red) color image. The middle and right of the bottom penal present the emission-line fitting in the \hb-\oiii\ region and the \ha-\nii\ region, respectively. The black and orange lines are the observed spectra and the model star light fitted with {\tt pPXF}, respectively. The red, blue, and green lines represent the synthetic spectra, models of narrow and wide components, and models of broad components, respectively.}
    \label{fig:fitting_result_example} 
\end{figure*}

In Figure~\ref{fig:fitting_result_example}, we present examples of the fitting continuum, SDSS {\it gri} color images, and emission-line fitting for sample \textbf{A}, sample \textbf{B}, and sample \textbf{C}. 

\begin{figure*}[thb]
    \centering
    \includegraphics[width=0.45\textwidth]{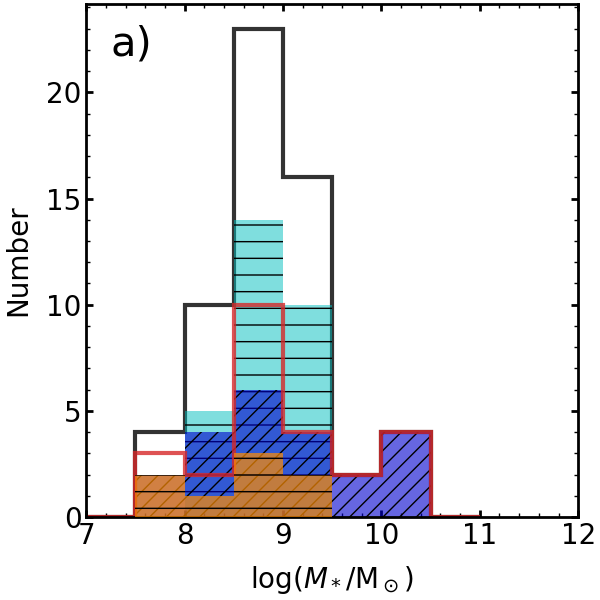}
    \includegraphics[width=0.45\textwidth]{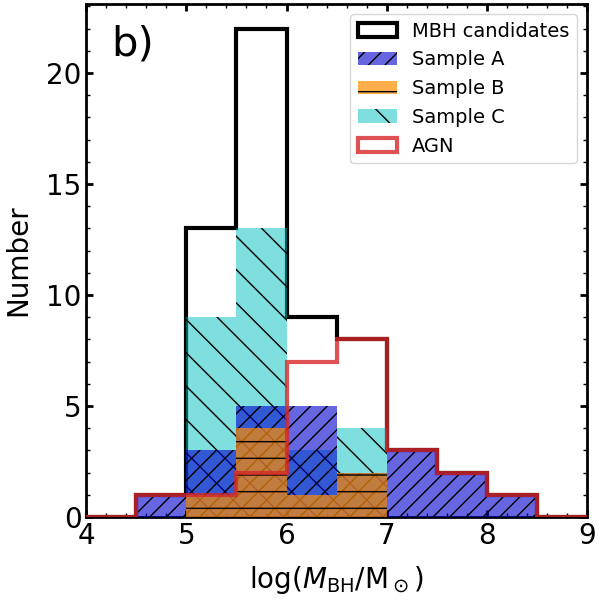}
    \caption{a) Distribution of stellar masses $\rm M_*$ for MBH candidates in GP galaxies. b) Distribution of BH masses ${\rm M_{BH}}$ for MBH candidates in GP galaxies. The black histogram illustrates the total sample of MBH candidates. The blue, orange, and cyan histograms show the sample \textbf{A}, sample \textbf{B}, and sample \textbf{C}, respectively. The red histogram represents the identified AGNs in this sample of MBH candidates.}
    \label{fig:his_mass}
\end{figure*}

\section{Estimation of Stellar Mass and BH Mass} \label{sec:mass_mea}
\subsection{Stellar mass}
\label{sec:stellar_mass}
We derive the stellar masses of broad-line candidates by modeling the Spectral Energy Distributions (SEDs) with {\tt CIGALE} \cite{Burgarella2005, Noll2009, Boquien2019}.

We assume a delayed star-forming history with five types of e-folding time ($\tau_{\rm main}$) and age ($t_{\rm main}$) of the main stellar population, which both range from 100 to 12000 Myrs. We adopt the Bruzual-Charlot Stellar Population Synthesis ({\tt BC03} SPS) \cite{Bruzual2003} with the Chabrier initial mass function (IMF) \cite{Chabrier2003} and 3 stellar metallicity (0.02, 0.2, and 0.4 Z$\odot$). 

We add the CLOUDY emission-line temples \cite{Inoue2011} to model the nebular lines, where the ionization parameter $\log U$ is set to -2.0. We employ the Charlot attenuation law \cite{Charlot2000} for gas and stellar attenuation and Dale templates \cite{ Dale2001, Dale2014} for dust emission. Since the AGN fraction may be high in our broad-line candidates, we also add the Stalevski AGN model \cite{Stalevski2016} and the AGN fraction is set from 0 to 0.5. Eventually, these set-up parameters lead to a grid of 10350000 models.

The input data is composed of {\it GALEX} NUV and FUV bands \cite{Morrissey2007}, SDSS {\it ugriz} {\tt cModel} photometry, and {\it WISE} W1, W2, W3, and W4 bands \cite{Wright2010}. We note that about half (27/59) of galaxies have {\it WISE} detection and most (44/59) of them have {\it GALEX} detection. We examine the impact of excluding {\it WISE} data on stellar mass measurements, finding an average deviation of -0.016 dex when {\it WISE} data are omitted, which can be negligible. 

Finally, the stellar mass \Mstar\ is estimated through the Bayesian approach. Our MBH candidates have stellar masses \Mstar\ in the range of $10^{7.61}$ to $10^{10.26}\,$\Msun\ with a median of $10^{8.83\pm0.11}\,$\Msun\ (see Figure~\ref{fig:his_mass} Panel a).

\subsection{BH Mass}
Analyzing the width and flux of the broad Balmer emission lines (e.g. \hb\ and \ha) is an efficient way to estimate the BH masses in low-mass galaxies, i.e. single-epoch virial BH masses \cite{Greene2005}. This involves inferring the BH mass from the relation \MBH$\ \propto R_{\rm BLR} \Delta V^2/G$, with parameters determined from the BLR velocity dispersion $\Delta V$, and the radius-luminosity relation between the luminosity $L$ of broad emission lines and the broad-line region (BLR) radius $R_{\rm BLR}$. A common scaling factor $\epsilon$ is applied to calibrate this reverberation-based BH masses to the \MBH-$\sigma_*$ relation. 

We estimate the BH masses following the equation (1) in ref. \cite{Reines2015} (the same equation (5) in  ref. \cite{Reines2013}): 
\begin{equation}
    \begin{split}
         \log(M_{\rm BH}/{\rm M_\odot}) &= \log\epsilon + 6.57 \\
        &+ 0.47\, \log(L_{\rm H\alpha^B}/10^{42}\rm\, erg\,s^{-1})\\
        & + 2.06\, \log(FWHM_{\rm H\alpha^B}/\rm 10^{3}\,km\,s^{-1}),
    \end{split}
\end{equation}
where $\epsilon$ = 1.075. The errors of BH masses are derived from the emission-line errors. We note that this estimation of BH mass has a typical uncertainty of $\sim 0.5$ dex (ref. \cite{Shen2013}, also see ref. \cite{Reines2015}). Finally, our MBH candidates have BH masses \MBH\ in the range of $10^{4.70}$ to $10^{8.47}\,$\Msun\ with a median stellar mass of $10^{5.85\pm0.64}\,$\Msun\ (see Figure~\ref{fig:his_mass} Panel b).

In our sample, the majority (36 out of 59) of MBH candidates are IMBH candidates with \MBH$\ < 10^6\,$\Msun\ (see Table~\ref{tab:parameterA} for their properties and Table~\ref{tab:agn} for the statistics).
We find only one IMBH candidate (J131253+171231) in our sample with \MBH$\ < 10^{5}\,$\Msun. 

\begin{table*} 
\centering 
\caption{Measured parameters of MBH candidates.} 
\label{tab:parameterA} 
\setlength{\tabcolsep}{0.2mm}{ 
\begin{tabular}{cccccccccccccc}  
\hline \hline 
name & {\it z}  & $L_{\rm H\alpha^B}$ & $L_{\rm H\alpha^N}$ & log $M_{\rm BH}$ & log $M_{*}$ &  $L_{X,b}$ & BL & BPT  & Mid-IR& X-ray& Radio & subset & dataset \\ 
 &   & $\rm 10^{41}erg\,s^{-1}$  &  $\rm 10^{41}erg\,s^{-1}$ &   &   & $\rm 10^{41} erg\,s^{-1}$ & AGN? & type & AGN? & AGN? & AGN? &   & \\
 (1) & (2) & (3) & (4) & (5) & (6) & (7) & (8) & (9) & (10) & (11) & (12) & (13) & (14)\\
\hline
J000430-101129 & 0.239 & 2.34(0.38) & 11.71(0.24) & 5.97(0.08) & 8.53(0.23) & -- & 0 & SF & 0 & 0 & 0 & B & sdss \\ 
J001523+292954 & 0.308 & 0.56(0.25) & 3.18(0.15) & 5.14(0.20) & 8.74(0.38) & -- & 0 & SF & 0 & 0 & 0 & C & sdss \\ 
J003515+084859 & 0.242 & 4.51(0.80) & 16.38(0.32) & 6.60(0.12) & 8.71(0.03) & -- & 1 & SF & 1 & 0 & 0 & C & lamost \\ 
J004913+002401 & 0.159 & 0.29(0.06) & 3.57(0.09) & 5.85(0.14) & 8.25(0.19) & -- & 0 & SF-Seyfert & 0 & 0 & 0 & A & sdss \\ 
J005504+250646 & 0.084 & 0.32(0.02) & 1.77(0.01) & 5.63(0.04) & 8.03(0.18) & -- & 0 & SF & 0 & 0 & 0 & B & lamost \\ 
J012606+281642 & 0.217 & 1.91(0.73) & 3.50(0.29) & 5.87(0.16) & 9.45(0.18) & -- & 0 & SF & 0 & 0 & 0 & C & lamost \\ 
J014846+142630 & 0.340 & 2.65(1.38) & 13.61(0.81) & 5.35(0.18) & 8.92(0.05) & -- & 0 & SF-Seyfert & 0 & 0 & 0 & C & sdss \\ 
J015013+105623 & 0.275 & 0.44(0.11) & 4.65(0.16) & 5.39(0.15) & 8.93(0.28) & -- & 0 & AGN & 0 & 0 & 0 & A & sdss \\ 
J020056+234517 & 0.244 & 0.71(0.16) & 6.25(0.16) & 5.86(0.13) & 8.87(0.31) & -- & 0 & Composite & 0 & 0 & 0 & A & sdss \\ 
J025055-043300 & 0.272 & 0.68(0.39) & 5.95(0.27) & 5.31(0.21) & 8.25(0.11) & -- & 0 & SF & 0 & 0 & 0 & C & sdss \\ 
J031623+000912 & 0.203 & 0.54(0.18) & 10.75(0.16) & 5.88(0.21) & 8.68(0.10) & -- & 0 & SF & 0 & 0 & 0 & A & sdss \\ 
J032613-063512 & 0.162 & 0.69(0.09) & 10.47(0.13) & 5.68(0.08) & 8.76(0.03) & -- & 0 & Composite & 1 & 0 & 0 & A & sdss \\ 
J081859+343238 & 0.340 & 7.86(0.59) & 15.75(0.45) & 6.22(0.04) & 9.34(0.08) & -- & 0 & Composite & 0 & 0 & 0 & A & sdss \\ 
J082927+250055 & 0.279 & 0.94(0.30) & 11.90(0.18) & 5.69(0.16) & 8.73(0.05) & -- & 0 & SF & 0 & 0 & 0 & C & lamost \\ 
J083833+374216 & 0.216 & 0.44(0.23) & 5.33(0.13) & 5.39(0.27) & 8.38(0.32) & -- & 0 & SF & 0 & 0 & 0 & C & sdss \\ 
J084029+470710 & 0.042 & 0.55(0.01) & 3.26(0.03) & 6.45(0.01) & 7.76(0.03) & $<0.8^a$ & 1 & SF & 0 & 0 & 0 & A & sdss \\ 
J084058+391925 & 0.249 & 0.65(0.20) & 5.16(0.16) & 5.42(0.14) & 9.06(0.30) & -- & 0 & SF & 0 & 0 & 0 & C & sdss \\ 
J085058+303053 & 0.280 & 0.78(0.23) & 4.46(0.15) & 6.33(0.19) & 8.33(0.08) & -- & 1 & SF-Seyfert & 0 & 0 & 0 & C & lamost \\ 
J085957+280123 & 0.295 & 1.85(0.44) & 3.82(0.23) & 5.94(0.14) & 8.84(0.22) & -- & 0 & SF-Seyfert & 0 & 0 & 0 & C & lamost \\ 
J091032+270719 & 0.282 & 1.51(0.44) & 9.04(0.28) & 5.98(0.17) & 8.79(0.06) & -- & 0 & SF & 0 & 0 & 0 & C & sdss \\ 
J091710+182949 & 0.312 & 1.14(0.28) & 11.16(0.21) & 5.71(0.13) & 9.43(0.26) & -- & 0 & SF-Seyfert & 0 & 0 & 0 & C & sdss \\ 
J092317+353452 & 0.269 & 1.77(0.49) & 6.09(0.21) & 5.68(0.14) & 9.01(0.22) & -- & 0 & SF & 0 & 0 & 0 & C & lamost \\ 
J092359+415736 & 0.323 & 6.64(0.46) & 6.22(0.19) & 6.92(0.04) & 10.03(0.20) & -- & 1 & AGN & 1 & 0 & 0 & A & sdss \\ 
J092438+470758 & 0.166 & 1.36(0.30) & 6.70(0.14) & 5.65(0.10) & 7.99(0.06) & -- & 0 & SF & 0 & 0 & 0 & B & lamost \\ 
J092834+292136 & 0.293 & 23.81(0.45) & 8.31(0.17) & 7.82(0.01) & 10.02(0.04) & -- & 1 & AGN & 1 & 0 & 0 & A & sdss \\ 
J093818+411740 & 0.280 & 3.36(1.10) & 8.63(0.73) & 6.53(0.23) & 8.83(0.04) & -- & 1 & SF-Seyfert & 0 & 0 & 0 & C & lamost \\ 
J093821-022334 & 0.185 & 1.67(0.16) & 9.46(0.08) & 6.62(0.07) & 8.67(0.02) & -- & 1 & SF & 1 & 0 & 0 & B & lamost \\ 
J094744+332556 & 0.302 & 0.79(0.21) & 8.26(0.18) & 5.84(0.13) & 8.91(0.03) & -- & 0 & SF & 1 & 0 & 0 & C & sdss \\ 
\hline 
\end{tabular} 
} 
    \begin{tablenotes} 
    \item NOTE. Parameters of MBH candidates. Column (1): name of galaxies. Column (2): spectroscopic redshift. Column (3)-(4): luminosity of the broad and narrow \ha\ components (in unit of $\rm 10^{41}\, erg\, s^{-1}$), respectively. We note that the luminosity of the wide component of \ha\ lines is not shown here. Column (5)-(6): log BH masses and stellar masses in \Msun, respectively. Column (7): X-ray luminosity in the full band. The symbols $<$ and -- represent non-detection and no cover, respectively. Column (8)-(12): classification via line widths, BPT diagram, Mid-IR color-color diagram, X-ray detection, and Radio excess to optical, respectively. 0 and 1 represent the true and false, respectively. Column (13): classes of subsets of MBH candidates, which is described in Sec.~\ref{sec:broad_criteria}. Column (14): spectrum dataset. The values in parentheses are 1$\sigma$ error.\\
    $^a$ The 2$\sigma$ detection limit of the {\it XMM-Newton} observation for this galaxy.\\
    $^b$ The 2$\sigma$ detection limit of the {\it Chandra} observation for individual galaxy.\\
    $^c$ The X-ray luminosity in the broad (0.5-7 keV) band with Chandra.\\
    \end{tablenotes} 
\end{table*}

\begin{table*} 
\setcounter{table}{0}
\centering 
\caption{Continued.} 
\label{tab:parameterB} 
\setlength{\tabcolsep}{0.2mm}{ 
\begin{tabular}{cccccccccccccc}  
\hline \hline 
name & {\it z}  & $L_{\rm H\alpha^B}$ & $L_{\rm H\alpha^N}$ & log $M_{\rm BH}$ & log $M_{*}$ &  $L_{X,b}$ & BL & BPT  & Mid-IR& X-ray& Radio & subset & dataset \\ 
 &   & $\rm 10^{41}erg\,s^{-1}$  &  $\rm 10^{41}erg\,s^{-1}$ &   &   & $\rm 10^{41} erg\,s^{-1}$ & AGN? & type & AGN? & AGN? & AGN? &  & \\
 (1) & (2) & (3) & (4) & (5) & (6) & (7) & (8) & (9) & (10) & (11) & (12) & (13) & (14)\\
\hline 

J095618+430727 & 0.276 & 3.98(0.60) & 13.17(0.23) & 6.75(0.07) & 9.07(0.02) & -- & 1 & SF & 0 & 0 & 0 & B & lamost \\ 
J102231+382804 & 0.268 & 1.23(0.62) & 8.31(0.36) & 5.20(0.20) & 8.28(0.09) & $<15.4^b$ & 0 & SF & 0 & 0 & 0 & C & sdss \\ 
J105114-005735 & 0.261 & 1.83(0.52) & 9.52(0.25) & 5.83(0.14) & 9.15(0.10) & -- & 0 & SF & 0 & 0 & 0 & C & lamost \\ 
J112615+385817 & 0.337 & 8.98(1.30) & 3.12(0.21) & 7.23(0.07) & 10.26(0.13) & -- & 1 & AGN & 1 & 0 & 1 & A & sdss \\ 
J114840+175633 & 0.079 & 0.35(0.02) & 7.28(0.06) & 6.29(0.03) & 8.39(0.10) & -- & 1 & SF-Seyfert & 0 & 0 & 0 & A & sdss \\ 
J115438+065025 & 0.269 & 14.79(0.68) & 3.39(0.17) & 7.68(0.03) & 10.08(0.10) & -- & 1 & AGN & 1 & 0 & 0 & A & lamost \\ 
J120052+331238 & 0.303 & 10.43(0.50) & 10.17(0.19) & 8.47(0.03) & 9.90(0.03) & -- & 1 & AGN & 1 & 0 & 0 & A & sdss \\ 
J122245+360218 & 0.301 & 4.73(0.72) & 10.97(0.31) & 6.74(0.11) & 8.62(0.13) & $<14.6^b$ & 1 & SF-Seyfert & 1 & 0 & 0 & A & sdss \\ 
J123723+390809 & 0.278 & 2.46(0.56) & 21.84(0.48) & 6.02(0.10) & 9.18(0.02) & -- & 0 & Composite & 1 & 0 & 0 & A & sdss \\ 
J124330-024241 & 0.210 & 0.79(0.11) & 4.21(0.03) & 6.61(0.08) & 8.62(0.34) & $<7.3^b$ & 1 & SF & 0 & 0 & 0 & C & lamost \\ 
J131253+171231 & 0.052 & 0.03(0.01) & 1.33(0.01) & 4.70(0.12) & 8.51(0.07) & -- & 0 & SF & 1 & 0 & 0 & A & sdss \\ 
J132103+363218 & 0.235 & 0.48(0.30) & 13.07(0.17) & 5.63(0.33) & 9.07(0.21) & -- & 0 & SF & 0 & 0 & 0 & C & lamost \\ 
J140551+515517 & 0.271 & 3.68(2.39) & 6.75(0.93) & 6.54(0.37) & 9.00(0.03) & -- & 1 & AGN & 1 & 0 & 1 & C & lamost \\ 
J141059+430246 & 0.066 & 0.09(0.03) & 5.28(0.05) & 5.11(0.21) & 8.17(0.04) & 0.3$^c$ & 0 & SF & 0 & 0 & 0 & A & sdss \\ 
J141550+182649 & 0.112 & 0.14(0.06) & 2.30(0.03) & 5.35(0.24) & 9.04(0.13) & -- & 0 & SF-Seyfert & 0 & 0 & 0 & B & lamost \\ 
J142644+271151 & 0.356 & 17.61(2.09) & 15.93(0.83) & 7.20(0.04) & 9.32(0.29) & -- & 1 & AGN & 0 & 0 & 1 & A & sdss \\ 
J144205-005248 & 0.282 & 1.67(0.71) & 16.65(0.46) & 5.51(0.20) & 9.20(0.09) & -- & 0 & SF-Seyfert & 0 & 0 & 0 & A & sdss \\ 
J145842+540852 & 0.183 & 0.61(0.12) & 4.92(0.06) & 5.52(0.10) & 8.93(0.13) & -- & 0 & SF-Seyfert & 0 & 0 & 0 & B & lamost \\ 
J152156+200224 & 0.323 & 1.12(0.37) & 11.92(0.31) & 5.57(0.15) & 9.02(0.19) & -- & 0 & SF & 0 & 0 & 0 & C & sdss \\ 
J152332+293112 & 0.068 & 0.20(0.02) & 4.60(0.05) & 5.35(0.04) & 8.03(0.13) & -- & 0 & SF & 0 & 0 & 0 & C & sdss \\ 
J152949+093144 & 0.303 & 0.82(0.34) & 6.75(0.26) & 5.67(0.24) & 8.99(0.28) & -- & 0 & SF-Seyfert & 0 & 0 & 0 & C & lamost \\ 
J154108+032029 & 0.310 & 2.03(1.05) & 17.68(0.58) & 5.38(0.20) & 8.82(0.16) & -- & 0 & SF-Seyfert & 0 & 0 & 0 & C & sdss \\ 
J155450+132609 & 0.149 & 0.13(0.04) & 4.86(0.08) & 5.32(0.21) & 8.33(0.03) & -- & 0 & SF-Seyfert & 0 & 0 & 0 & A & sdss \\ 
J160550+440540 & 0.320 & 8.87(0.66) & 11.41(0.24) & 7.23(0.03) & 9.96(0.08) & 180.8$^c$ & 1 & AGN & 1 & 1 & 1 & A & sdss \\ 
J165129+254250 & 0.237 & 1.07(0.45) & 4.03(0.25) & 5.35(0.17) & 9.17(0.25) & -- & 0 & SF & 0 & 0 & 0 & C & sdss \\ 
J172932+315219 & 0.310 & 0.83(0.40) & 7.09(0.22) & 5.51(0.20) & 8.83(0.13) & -- & 0 & SF & 0 & 0 & 0 & C & sdss \\ 
J222803+162200 & 0.226 & 0.58(0.14) & 6.42(0.07) & 6.28(0.15) & 8.56(0.24) & -- & 1 & SF-Seyfert & 0 & 0 & 0 & C & lamost \\ 
J225059+000032 & 0.081 & 0.13(0.01) & 1.43(0.02) & 6.21(0.05) & 7.96(0.10) & -- & 1 & SF-Seyfert & 0 & 0 & 0 & A & sdss \\ 
J225108-003013 & 0.081 & 0.23(0.02) & 2.51(0.01) & 6.19(0.05) & 7.61(0.31) & -- & 1 & SF & 0 & 0 & 0 & B & lamost \\ 
J232133+174029 & 0.303 & 1.73(0.81) & 18.72(0.37) & 6.07(0.23) & 9.25(0.05) & -- & 0 & Composite & 0 & 0 & 0 & C & lamost \\ 
J233446+202530 & 0.301 & 0.41(0.17) & 4.82(0.13) & 5.61(0.21) & 8.74(0.13) & -- & 0 & SF-Seyfert & 0 & 0 & 0 & C & sdss \\ 
\hline 
\end{tabular} 
} 
\end{table*}

\begin{figure*}[htb]
    \centering
    \includegraphics[width=0.35\textwidth]{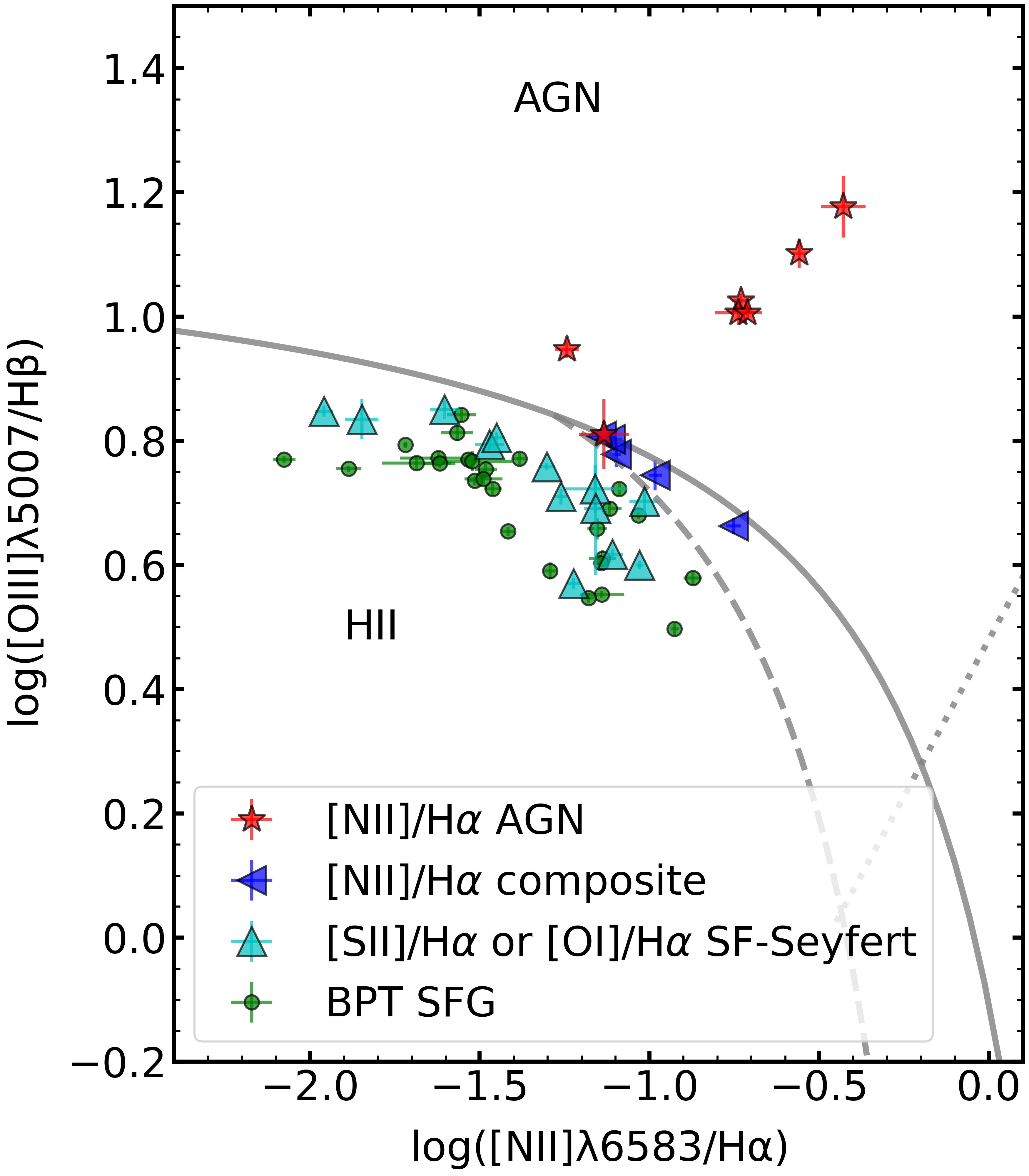}
    \includegraphics[width=0.3\textwidth]{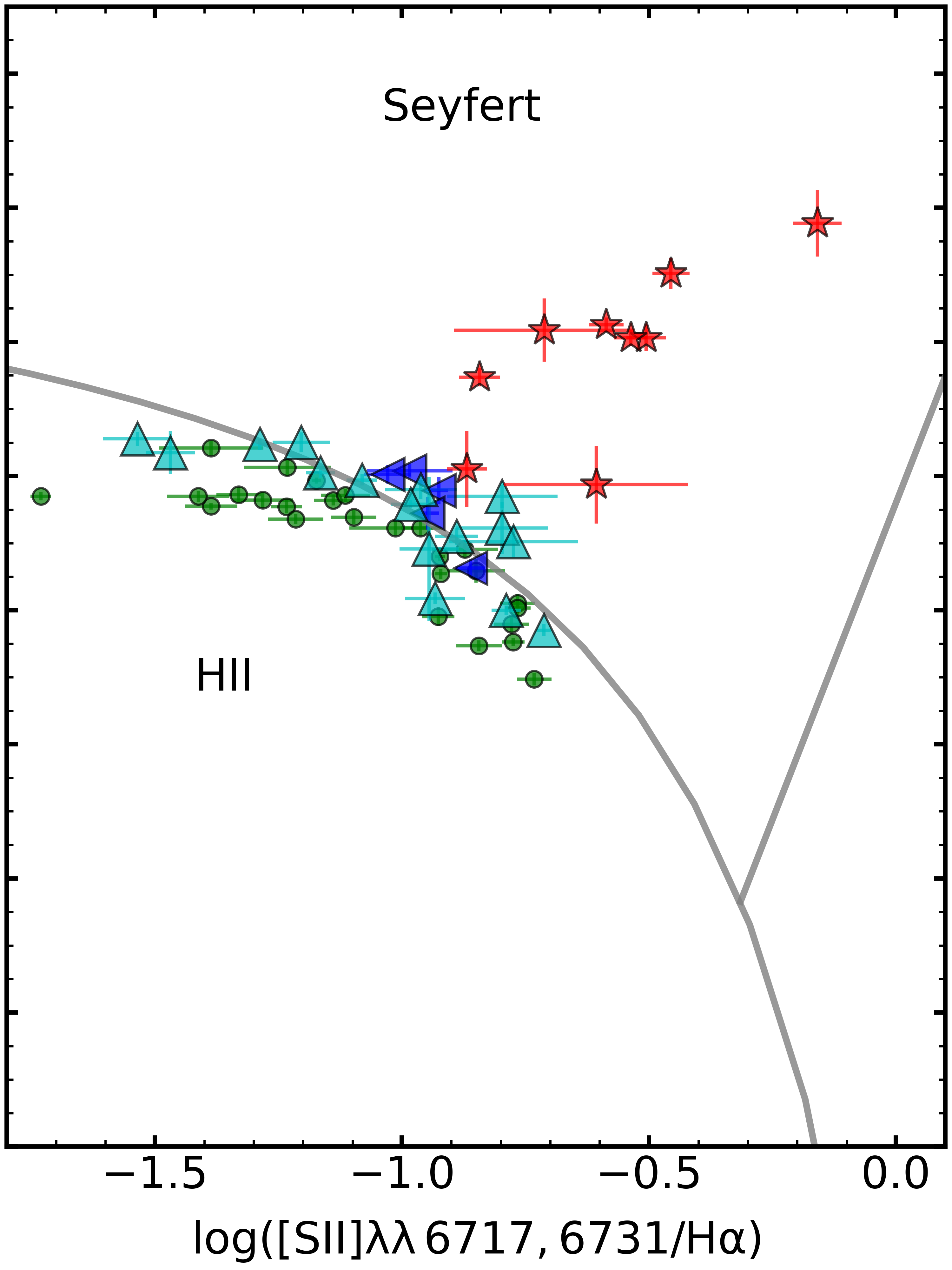}
    \includegraphics[width=0.3\textwidth]{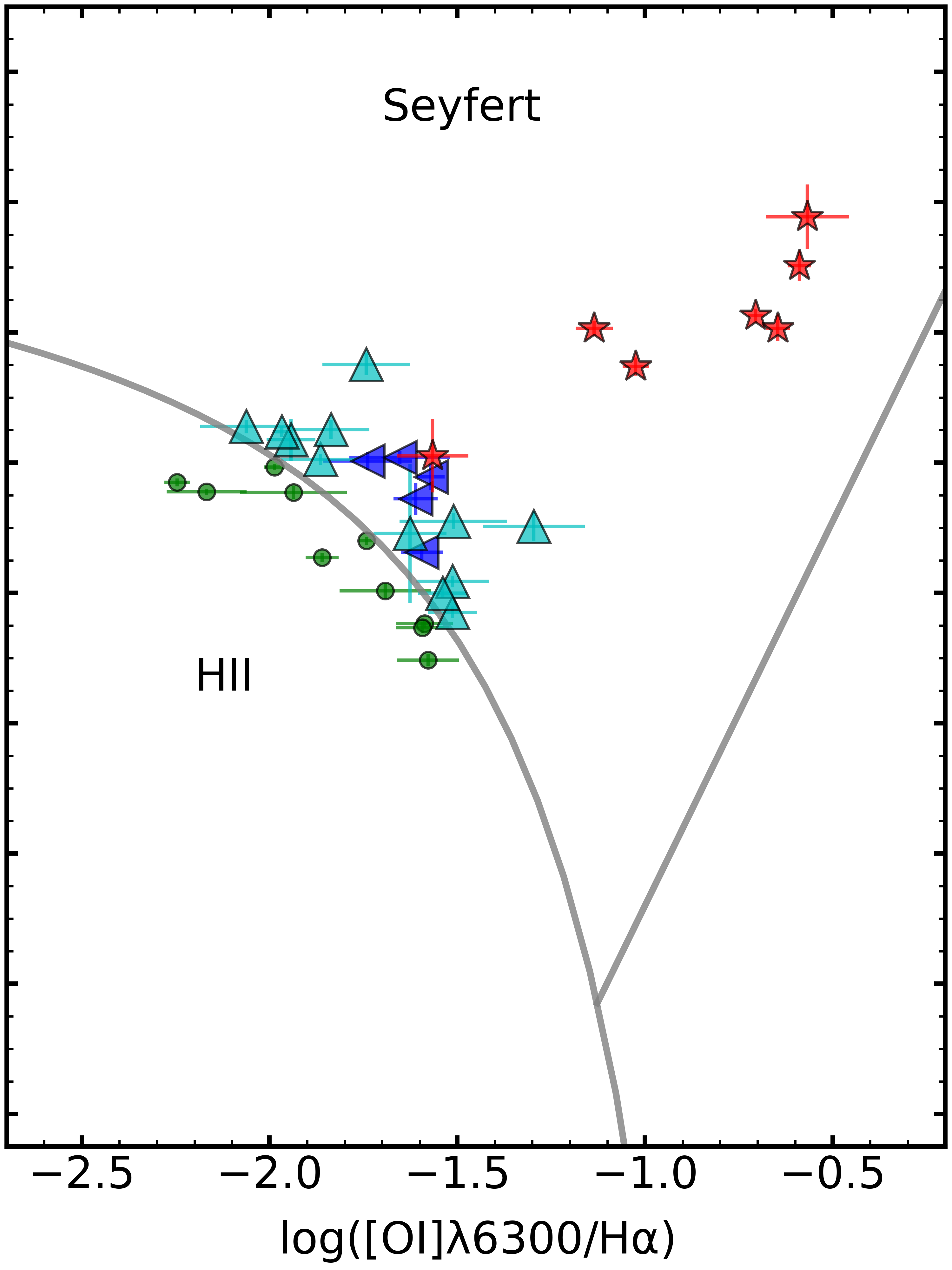}
    \caption{BPT diagram for MBH candidates in GP galaxies. Left: The \oiii/\hb\ vs. \nii/\ha\ diagram; Middle: The \oiii/\hb\ vs. \sii/\ha\ diagram; Right: The \oiii/\hb\ vs. \oi/\ha\ diagram. In these three diagrams, only galaxies that have S/N$_{\rm [NII]}$, S/N$_{\rm [SII]}$, and S/N$_{\rm [OI]} > 3$ are plotted, respectively. We also include sources with flux ratios of \oiii/\hb\ $>$ 10 in these diagrams. The classiﬁcation lines follow \cite{Kewley2001} (solid line) and \cite{Kauffmann2003} (dashed and dotted lines), also seen in \cite{Kewley2006}. Red stars and blue triangles represent the AGNs and composite galaxies classified by the \nii-BPT diagram, respectively. The SF-Seyfert galaxies classified on the \sii-BPT or \oi-BPT diagram are colored in Cyan. The star-forming galaxies are marked in green dots.
    }
    \label{fig:bpt}
\end{figure*}


\begin{table*}
\centering
\caption{Statistics of identified AGN and IMBH candidates.}
\label{tab:agn}
\setlength{\tabcolsep}{4mm}{
\begin{tabular}{cccccccc} 
\hline \hline
\# & Broad-line AGN  & BPT AGN  & Mid-IR AGN  & X-ray AGN  & Radio AGN  & IMBH  \\ 
 & (\#59) & (\#53) & (\#27) & (\#6) & (\#59) & \\
 & (1) & (2) & (3) & (4) & (5) & (6) \\
\hline
Sample A (\#22) & 11 & 8 & 10 & 1 & 3 & 9  \\
Sample B (\#8)  & 3  & 0 & 1  & 0 & 0 & 5  \\
Sample C (\#29) & 6  & 1 & 3  & 0 & 1 & 22  \\
total (\#59)    & 20 & 9 & 14 & 1 & 4 & 36  \\
\hline
\end{tabular}
}
    \begin{tablenotes}
    \item NOTE. Column (1)-(5): Numbers of AGN identified as broad-line AGN, BPT AGN (in the \nii-BPT diagram, specifically), Mid-IR AGN,  X-ray AGN, and Radio AGN respectively. The numbers in parentheses indicate the source counts covered by the respective datasets (see Section \ref{sec:AGN_id} in detail). Column (6): IMBH candidates.
    \end{tablenotes}
\end{table*}

\section{AGN Identification} \label{sec:AGN_id}
\subsection{Multi-band Information}
To find other evidence of existing active BHs in these galaxies, we employ various methods, including the broad-line width, the optical narrow emission-line diagnostic diagrams (i.e. the BPT diagram), the mid-infrared color-color diagram, the X-ray emission, and the radio emission. The results of the various identifying methods are listed below. However, we should note that all these methods are well-tested with the activities of SMBHs, while very little is known about the activities of IMBHs. The limits of these methods on IMBHs are discussed in the next subsection.

\begin{itemize}
    \item {\it \ha\ line width.} There are 20 out of 59 MBH candidates having broad \ha\ component widths of $FWHM_{\rm H\alpha}^B > 1000\rm\, km\, s^{-1}$, suggesting the presence of the BLR. Therefore, we consider these candidates as broad-line AGNs.
    
    \item {\it The BPT classification.} The BPT diagram is an efficient method to separate star-forming galaxies and AGNs based on the hardness of narrow lines. In Figure~\ref{fig:bpt}, 
    we plot \oiii/\hb\ as a function of \nii/\ha, \sii/\ha\ and \oi/\ha\ (hereafter \nii-, \sii-, and \oi-BPT diagrams), respectively, 
    where we require the sources with S/N$_{\rm [NII]}$, S/N$_{\rm [SII]}$, and S/N$_{\rm [OI]}$ $>3$ for each corresponding diagram, respectively. 
    With the classification criteria of ref. \cite{Kewley2001, Kauffmann2003, Kewley2006} and the criteria of \oiii/\hb\ $>$ 10, we classify 9 AGNs, 5 composite galaxies, 18 SF-Seyfert galaxies, and 27 star-forming galaxies in our MBH candidates. Here the SF-Seyfert galaxies are those galaxies classified as Seyfert galaxies only in the \sii-\ or \oi-BPT diagram, which are mainly low-metallicity AGNs or/and Seyfert galaxies dominated by star formation (the same definition of SF-AGN in ref. \cite{Polimera2022}). Since the AGN contribution of SF-Seyfert galaxies is only 8-16\% \cite{Polimera2022}, we are not including these galaxies in the identified AGN sample.
    
    \item {\it Mid-IR Color.} We collect the mid-infrared information from the \href{https://wise2.ipac.caltech.edu/docs/release/allwise/}{ALLWISE source catalog}, which is an imaging survey covering the entire sky.
    Over half (27/59) of MBH candidates show mid-infrared detection with S/N $>$ 2 in both W1, W2, and W3 bands. To search for possible mid-infrared AGNs, we follow the color-color criteria in ref. \cite{Jarrett2011} and present the color-color distribution in Figure~\ref{fig:wise}. 
    Among the 27 galaxies with mid-infrared data, 14 meet the color criteria and hence are identified as mid-infrared AGNs.

\begin{figure}[H]
    \centering
    \includegraphics[width=0.48\textwidth]{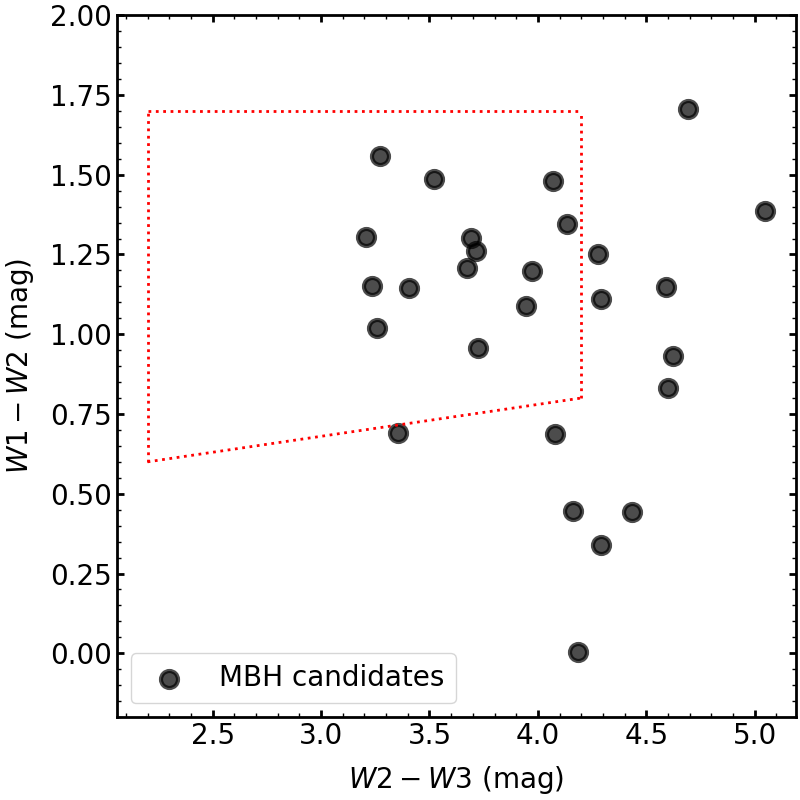}
    \caption{{\it WISE} color-color diagram for MBH candidates in GP galaxies. Only galaxies that have detections with S/N $>2$ in W1, W2, and W3 bands are plotted. The red dashed lines mark the mid-infrared AGN selection criteria from ref. \cite{Jarrett2011}.}
    \label{fig:wise}
\end{figure}

    \item {\it X-ray Emission.} We collect the available X-ray photometry from both the {\it Chandra} Source Catalog (\href{https://cxc.cfa.harvard.edu/csc}{CSC 2.0} \cite{Evans2010, Evans2020}) and the {\it XMM-Newton} serendipitous source catalog (\href{http://xmmssc.irap.omp.eu/Catalogue/4XMM-DR11/4XMM_DR11.html}{4XMM-DR9} \cite{Webb2020}) for our MBH candidates. The CSC covers $\rm \sim 560\ deg^2$ of the sky and reaches the deepest detection of $L_{\rm{FB}}=10^{38.5}\,\rm{erg\,s^{-1}}$ in the full (0.5-7 keV) band at $z$ $\sim$ 0.3. The 4XMM-DR9 covers 1152 $\rm deg^2$ of the sky and have a median luminosity of $L_{\rm{FB}}=10^{42.8}\,\rm{erg\,s^{-1}}$ in the total (0.2-12 keV) band \cite{Rosen2016} at $z$ $\sim$ 0.3. But only 1 and 5 MBH candidates are covered by {\it XMM-Newton} and {\it Chandra}, respectively. Among these 6 candidates, only two galaxies (J141059+430246 and J160550+440540) have a {\it Chandra} detection of $L_{\rm FB} = 3\times10^{40}$ and $2\times10^{43}\,\rm erg\,s^{-1}$, respectively. 
    While the X-ray luminosity of the former can be explained by the star-forming activities, that of the latter is too intense to originate from the star-forming region. Therefore, we speculate that J160550+44054 is an AGN. 
   In Table~\ref{tab:parameterA}, we give the detection limit at 2$\sigma$ of {\it XMM-Newton} or {\it Chandra} observations in the nearby region (within 10$\arcmin$) for four galaxies with non-detection.
    Since X-ray detection is a potent AGN indicator, the absence of X-ray observations in our sample, coupled with weak X-ray emission from IMBHs, poses challenges to reliably identifying AGNs among these MBH candidates.

    \item {\it Radio Emission.} The Faint Images of the Radio Sky at Twenty-cm (\href{http://first.astro.columbia.edu}{FIRST} \cite{Becker1995}) survey used the NRAO Very Large Array (VLA) to explore the radio radiation of optical sample at $\sim$ 1.4 GHz, covering over 10000 deg$^2$ and reaching a depth of 0.15 mJy. We explore the radio properties of this MBH sample with FIRST. There are 4 out of 59 candidates with radio detection of $f_{\rm 1.4GHz}$ ranging from 0.87 to 2.21 $\rm mJy$ (with a typical rms noise of $\rm \sim0.19\ mJy$), corresponding to a radio star formation rate SFR$_\textrm{radio}$ in the range of 126-470 $\rm M_\odot\,yr^{-1}$ (with the conversion from Equation (17) in ref. \cite{Murphy2011}), respectively. These SFRs have an excess over those derived from the narrow \ha\ luminosity ($\rm SFR_{H\alpha} \sim$ 22-116\, $\rm M_\odot\,yr^{-1}$), with the conversion from ref. \cite{Kennicutt2012}, suggesting the excess radio emission is produced from AGN. Therefore, we classify these 4 MBH candidates as radio AGNs. 
\end{itemize}

\subsection{Final Classification and Limitations}
The final classification is compiled in Table~\ref{tab:agn}. Eventually, there are 25 out of 59 MBH candidates classified as AGNs. 

However, these different AGN-identifying methods have certain limitations. The broad-line width method ignores MBHs with narrower line widths. The BPT diagram may miss some AGNs in low-metallicity galaxies, and GP galaxies usually have relatively low metallicity \cite{Amorin2010, Izotov2011}, so the AGN candidates missed by this method are not negligible. The mid-infrared color diagram is commonly used to search for obscured AGNs, and in our sample, it is relatively effective because of the larger coverage. Since star formation can also contribute to mid-infrared radiation, contamination from the star formation region is possible. As for the X-ray detection, our sources are less covered by the X-ray dataset, making it challenging to discuss the effectiveness of the X-ray detection in identifying IMBHs in our candidate sample. For the radio method, we assess the potential presence of jets by measuring radio emissions exceeding optical star formation rates. Yet, most sources do not exhibit strong radio emission, with only three sources detected. Additionally, since GPs have a high star formation rate, this method may not be effective in searching for AGNs in star-forming galaxies.
 
Moreover, as illustrated in Figure~\ref{fig:his_mass}, the successful identification of AGNs is correlated with BH mass with more massive galaxies identified more likely as AGNs. It seems that the traditional identification methods, typically applicable to SMBHs, may not effectively respond to the activity of IMBHs.

\section{Discussion} \label{sec:discussion}
\begin{figure*}[thbp]
    \centering
    \includegraphics[width=1\textwidth]{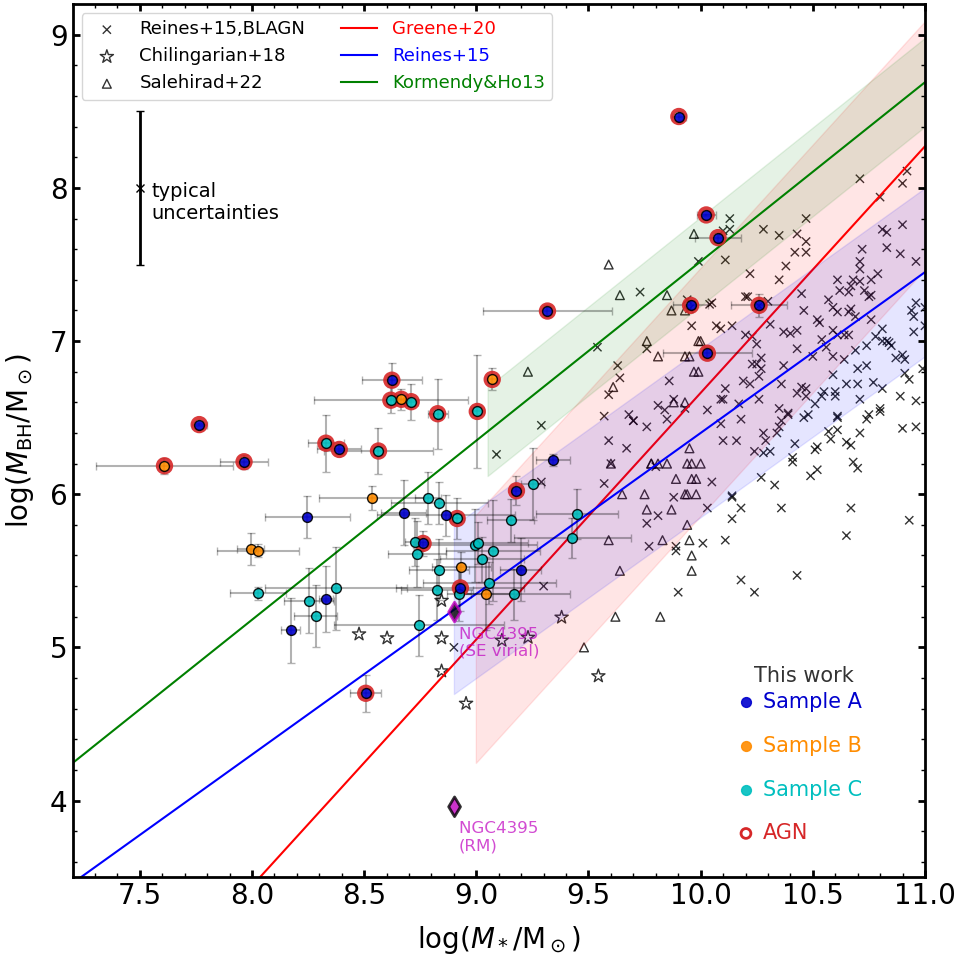}
    \caption{Single-epoch virial BH masses \MBH\ versus total stellar masses \Mstar\ for MBH candidates in the sample of GP galaxies. The blue, orange, and cyan dots are MBH candidates of sample \textbf{A}, sample \textbf{B}, and sample \textbf{C}, respectively. The identified AGNs are marked in red circles. Gray error bars indicate the estimated errors derived from emission-line measurements, and the black error bar in the upper left denotes the uncertainty of single-epoch virial BH masses. The blue and red solid lines illustrate the \MBH$-$\Mstar\ relation derived for dwarf galaxies \cite{Reines2015} and for more massive galaxies \cite{Greene2020}, respectively. The green solid line represents \MBH$-M_{\rm bulge}$ of SMBHs and their host galaxies \cite{Kormendy2013}. Shaded areas with corresponding colors show intrinsic scatter and the lower cuts indicate the lowest bulge mass or stellar mass in each relation. Samples of broad-line AGNs from ref. \cite{Reines2015} are shown as gray crosses. BH samples reported in ref. \cite{Chilingarian2018} and \cite{Salehirad2022} are plotted as gray stars and triangles, respectively. NGC4395, hosting the smallest BH known with a stellar mass of $10^{8.9}\,$\Msun, is also included, with BH masses determined using single-epoch (SE) virial and reverberation-mapping (RM) methods \cite{LaFranca2015, Woo2019}.}
    \label{fig:MBH_M}
\end{figure*}

\subsection{\MBH\ vs. \Mstar\ in GPs}
The scaling relations between BH mass and galaxy mass may encode the co-evolutionary history of galaxies and BHs. 
In Figure~\ref{fig:MBH_M}, we show the location of our MBH candidates in the \MBH $\ vs.$ \Mstar\ diagram. In general, our sample is consistent with the \MBH$-M_{\rm bulge}$ relation derived from SMBHs in classical bulges and elliptical galaxies (ref. \cite{Kormendy2013}, green line, $10^{9.05}\leq M_{\rm bulge}\leq 10^{12.09}$ \Msun), while shows large scatter at the low-mass end. 
Furthermore, our sample is above 
the \MBH$-$\Mstar\ relation derived from MBHs in dwarf galaxies (ref. \cite{Reines2015}, blue line, $10^{8.90}\leq M_*\leq 10^{11.49}$ \Msun) and from MBHs in more massive galaxies (ref. \cite{Greene2020}, red line, $10^{9.0}\leq M_*\leq 10^{11.9}$ \Msun).
Compared to the sample of ref. \cite{Reines2015}, our sample's median stellar mass and BH mass are 1.64 dex and 0.98 dex lower, respectively.

The portion of our sample excessing these relations is similar to broad-line AGNs with overmassive BHs at high redshift discovered by JWST recently \cite{Maiolino2023, Greene2023, Harish2023, Kocevski2023, Kokorev2023, Larson2023, Matthee2023} (Lin et al. in prep.). This is consistent with the model of BHs with super-Eddington accretion during gas-rich mergers \cite{Trinca2022}. It also indicates the fact that due to the evolutionary differences, the \MBH$-$\Mstar\ relations vary between different types of galaxies. Furthermore, we cannot rule out the possible scenario that leftover BH seeds in isolated dwarf galaxies have recently become active \cite{Mezcua2017}. 

In Figure~\ref{fig:MBH_M}, we also show the comparison of our MBH candidates with other IMBH samples.
Ref. \cite{Chilingarian2018} searched for IMBHs with broad \ha\ emission in the SDSS DR7 spectra using a non-parametric model. By identifying reliable AGNs through the BPT classification, X-ray detection, and broad \ha\ emission in second-epoch spectra, they reported a sample of 10 IMBHs with high confidence. Their sample is located below the \MBH$-$\Mstar\ relation. In contrast to their sample, our BH sample is more massive and above the \MBH$-$\Mstar\ relation.

Ref. \cite{Salehirad2022} presented a sample of black holes in the form of AGNs or TDEs with broad \ha\ emission from low-mass (\Mstar$\ < 10^{10}\,$\Msun) galaxies in the Galaxy and Mass Assembly survey. 
They utilized narrow emission-line ratios and high-ionization emission lines to confirm the presence of AGNs or TDEs and thus obtained a BH sample with a wide range of galaxy properties. Compared to their sample, our sample exhibits a similar range of BH masses but has a wider distribution of galaxy masses. The different SED fitting methods may induce this difference in galaxy masses.

We note that different methods have been used to estimate stellar masses, including the color-dependent mass-to-light ratio and the SED fitting methods with different parameters. Even considering the systematic offset of stellar masses led by various methods (up to $\sim$ 0.24 dex, see \ref{appendix} for details), our conclusion remains unchanged.

\subsection{How reliable are these IMBH candidates?}
Only 4 out of 36 IMBH candidates are identified as AGNs, either via the mid-IR color-color diagram or BPT diagram, indicating reliable BH signals. Additionally, there are 12 IMBH candidates classified as composite or SF-Sayfert galaxies, which may have a lower contribution of AGNs.
However, the remaining 20 IMBH candidates (2 in sample \textbf{A}, 3 in sample \textbf{B}, and 15 in sample \textbf{C}) have BH evidence only in the broad \ha\ component without confirmation with other methods. Here we discuss the possible contaminations from SNe explosion and TDEs in the whole MBH sample below.

Transient stellar activities such as Type IIn SNe (SNe IIn) explosions and SNe Ia-CSM can also produce broad \ha\ emission with line widths up to several thousand $\rm km\,s^{-1}$ lasting a few years \cite{Taddia2013, Yang2023, Wang2023}, where SNe Ia-CSM is a rare type of Type Ia SNe with strong interaction between SNe Ia and their circumstellar medium (CSM). Moreover, due to the interaction between opaque CSM and ejecta of SN, the flux and line width of the \ha\ emission vary across time. One of the brightest SNe IIn known has a total \ha\ luminosity at peak of $\rm 10^{42}\, erg\,s^{-1}$ \cite{Fransson2014} and SNe Ia-CSM have similar luminosity of \ha\ to extreme cases of SNe IIn \cite{Yang2023, Wang2023}. Note that most sources in our sample have a broad \ha\ luminosity less than $\rm 10^{42}\, erg\,s^{-1}$, which probably originates from the explosion of a single SN IIn or SN Ia-CSM. 

Multiple-epoch spectroscopic observations or detections of variability can help exclude these transient events. 
Therefore, we analyze the spectra of candidates with second-epoch spectroscopic observations, using the same method as ref. \cite{Baldassare2016}. There are 6 sources (J003515+084859, J004913+002401, J084029+470710, J152156+200224, J160550+440540, J225059+000032) having valid second-epoch spectra. All of these sources exhibit broad \ha\ emission in their second-epoch spectra. Yet, only 3 of them (J004913+002401, J084029+470710, J225059+000032) have widths or fluxes consistent with the previous results within the margin of error. Besides transient broad \ha\ lines, the possible reason for those differences in line widths and fluxes could be shallower observation with LAMOST and the AGN variability. Although the longest time scale of two epochs is less than $\sim$ 1 year, which is not long enough to rule out the contribution of the SN IIn explosion, there is no detectable presence of P-Cygni profiles in these spectra, suggesting a low event rate of SNe IIn in this sample. 

We estimate the rate of SNe IIn and SNe Ia-CSM in our parent sample.
According to the SN rate-size relation for SNe II and SNe Ia in galaxies~\cite{Li2011}, the SN rate per unit mass {\it SNR} can be described as:
\begin{eqnarray}
 \log(\frac{\rm SNR(\textrm{SNe II/Ibc})}{10^{12}}) & = & -0.55\times \log(\frac{M_{*}}{10^{10}})-0.073,\\
\log(\frac{\rm SNR(\textrm{SNe Ia})}{10^{12}}) & = & -0.50\times \log(\frac{M_{*}}{10^{10}})-0.65,
\end{eqnarray}
where the {\it SNR} is in the unit of $\rm yr^{-1}\,M_{\odot}^{-1}$.
For our parent sample with a median stellar mass of $10^{8.8}\,$\Msun, the corresponding event rates of SNe II and SNe Ia are $\sim$ 0.001 $\rm yr^{-1}$ and $\sim$ 0.003 $\rm yr^{-1}$, respectively. Furthermore, the SNe IIn and SNe Ia-CSM are only $\sim$ 9\% and 0.2\% of the total SNe II and SNe Ia, respectively \cite{Smith2011, Sharma2023}. Thus, the event rate of SNe IIn and SNe Ia-CSM is $\sim$ 0.0001$\rm\, yr^{-1}$ and the expected number of SN IIn and SNe Ia-CSM in our parent sample is $\rm \sim 0.2\, yr^{-1}$. From this aspect, the contamination of SNe IIn and SNe Ia-CSM in our sample can be negligible.

However, due to the lack of multi-epoch spectral observations for all sources, we cannot completely rule out the contribution of SN IIn explosion in the sample of MBH candidates with current data. Confirming reliable BHs in our samples requires deeper follow-up spectroscopic observation over a longer time scale or deep X-ray observations.

TDEs can also induce \ha\ emission line reaching $\rm \sim 10^{42}\, erg\,s^{-1}$ at peak \cite{Charalampopoulos2022}, directly from TDEs or the interaction between TDEs and surrounding gas. The \ha\ line widths derived from these two mechanisms are different and vary from thousands to $\rm 10^{4}\, km\,s^{-1}$ \cite{vanVelzen2020, Gezari2021, Wang2022}. The TDE rate is $\rm \sim 10^{-5}\, yr^{-1}\, gal^{-1}$ \cite{Stone2016}. Therefore, there are  $\rm \sim 0.02\, yr^{-1}$ TDEs expected in our parent sample, indicating that the contamination of TDEs in our sample can be also  negligible.

In summary, it is very unlikely that the identified AGNs (25 out of 59) in our MBH candidates are contaminated by SNe explosions. For the remaining sources that are not identified as AGNs yet, follow-up observations are required to determine the origin of the broad Ha emissions.

\section{Conclusion} \label{sec:conclu}
Based on the broad \ha\ emissions in spectra, we find 59 MBH candidates in 2190 GP galaxies from LAMOST and SDSS surveys. We estimate the stellar masses and BH masses of this sample. Our main results are summarised as follows:  
\begin{itemize}
    \item Based on the profiles of \oiii\ $\lambda 5007$ lines, we decompose the Ha line into several components, and extract the broad Ha emission, leading to a new sample of 59 MBH candidates.
    \item For the MBH candidate sample, the stellar mass distribution ranges from $10^{7.61}$ to $10^{10.26}\,$\Msun\ and the BH mass distribution ranges from $10^{4.7}$ to $10^{8.5}\,$\Msun. Moreover, the majority (36/59) of MBH candidates are IMBH candidates with \MBH$\ < 10^6\,$\Msun\ and the smallest one in this sample has \MBH\ $\sim 10^{4.7}\,$\Msun. 
    Accounting for measurement errors, these MBH candidates are consistent with the \MBH$-M_{\rm bulge}$ relation for SMBHs in literature.
    \item To re-examine the AGN activities among this sample, we utilize various methods such as the broad-line width, BPT diagram, mid-infrared color, X-ray luminosity, and radio emission. There are 25 MBH candidates identified by at least one method. 
    We discuss the possible contamination by SNe IIn and TDEs and find only 0.2 $\rm yr^{-1}$ SNe IIn and 0.02 $\rm yr^{-1}$ TDEs expected in these MBH candidates. 
\end{itemize} 
The identified MBH sample provides an intriguing opportunity to probe the coevolution between MBHs and GP galaxies, which are believed to be the low-redshift analogs of high-redshift star-forming galaxies. Additionally, our work shows that it is very promising to find IMBHs in compact, star-forming dwarf galaxies.

\Acknowledgements{We thank Luis C. Ho, Tinggui Wang, Hengxiao Guo, and Haicheng Feng for the constructive discussions and comments.
Z.Y.Z. acknowledges the support of the National Key R\&D Program of China No.2022YFF0503402, the National Science Foundation of China (12022303), and the China-Chile Joint Research Fund (CCJRF No. 1906). L.Z.W is sponsored (in part) by the Chinese Academy of Sciences (CAS), through a grant to the CAS South America Center for Astronomy (CASSACA) in Santiago, Chile. J.X.W. acknowledges the support of the National Natural Science Foundation of China (NSFC) No.12033006 and No.12192221.
We acknowledge the science research grants from the China Manned Space Project with No. CMS-CSST-2021-A07, CMS-CSST-2021-A04 and CMS-CSST-2021-B04.
Guoshoujing Telescope (the Large Sky Area Multi-Object Fiber Spectroscopic Telescope LAMOST) is a National Major Scientific Project built by the Chinese Academy of Sciences. Funding for the project has been provided by the National Development and Reform Commission. LAMOST is operated and managed by the National Astronomical Observatories, Chinese Academy of Sciences.
Funding for the Sloan Digital Sky Survey V has been provided by the Alfred P. Sloan Foundation, the Heising-Simons Foundation, the National Science Foundation, and the Participating Institutions. SDSS acknowledges support and resources from the Center for High-Performance Computing at the University of Utah. The SDSS web site is \url{www.sdss.org}. SDSS is managed by the Astrophysical Research Consortium for the Participating Institutions of the SDSS Collaboration, including the Carnegie Institution for Science, Chilean National Time Allocation Committee (CNTAC) ratified researchers, the Gotham Participation Group, Harvard University, Heidelberg University, The Johns Hopkins University, L’Ecole polytechnique f{\'e}d{\'e}rale de Lausanne (EPFL), Leibniz-Institut f{\"u}r Astrophysik Potsdam (AIP), Max-Planck-Institut f{\"u}r Astronomie (MPIA Heidelberg), Max-Planck-Institut f{\"u}r Extraterrestrische Physik (MPE), Nanjing University, National Astronomical Observatories of China (NAOC), New Mexico State University, The Ohio State University, Pennsylvania State University, Smithsonian Astrophysical Observatory, Space Telescope Science Institute (STScI), the Stellar Astrophysics Participation Group, Universidad Nacional Aut{\'o}noma de M{\'e}xico, University of Arizona, University of Colorado Boulder, University of Illinois at Urbana-Champaign, University of Toronto, University of Utah, University of Virginia, Yale University, and Yunnan University.
This research has made use of data obtained from the Chandra Source Catalog, provided by the Chandra X-ray Center (CXC) as part of the Chandra Data Archive.
}

\InterestConflict{The authors declare that they have no conflict of interest.}



\bibliography{ref}{}
\bibliographystyle{scpma}

\begin{appendix}
\renewcommand{\thesection}{Appendix}
\section{Uncertainty of Stellar Mass Estimation\label{appendix}}
\begin{figure}[H]
    \centering
    \includegraphics[width=0.45\textwidth]{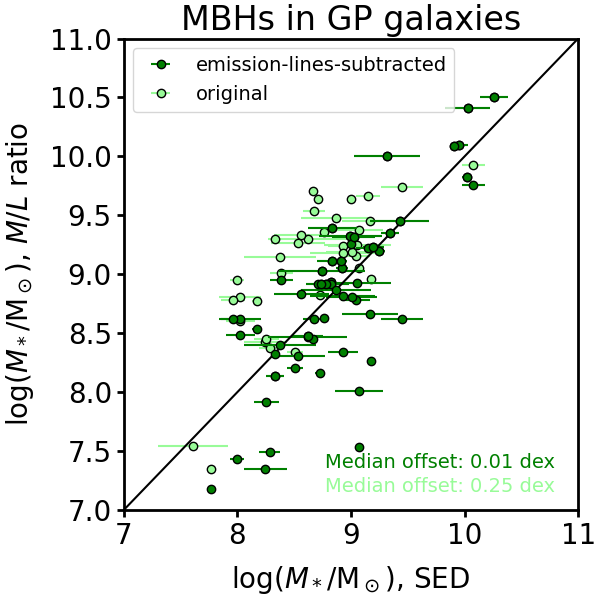}
    \caption{Comparison of the stellar mass of our MBH sample derived from the SED fitting and mass-to-light ratio. The light green and dark green dots represent stellar masses derived by the mass-to-light ratio with original and emission-line-subtracted {\tt cModel} broadband photometries, respectively. The median offsets between stellar masses estimated from the SED fitting and the mass-to-light ratio are also labeled in the lower right. The solid lines indicate the one-to-one relation.}
    \label{fig:massComparison2}
\end{figure}

We derive stellar masses of GP galaxies using SED fitting with UV-MIR data as described in Section \ref{sec:stellar_mass}. While, the stellar masses can also be estimated by the color-depended mass-to-light ratios, which are well used in host galaxies of AGN at low-redshift in some works (e.g., refs \cite{Reines2015, Dong2012, Chilingarian2018}). In contrast to those works, our sample has significant emission lines (e.g., \oiii\ and \ha), which affect the broadband photometry in some bands, leading to an overestimation ($\sim 0.25$ dex; see Figure~\ref{fig:massComparison2}) of the stellar masses when deriving the stellar masses based on the mass-to-light ratios. In Figure~\ref{fig:massComparison2}, we compare the stellar masses derived from the SED fitting (see Section~\ref{sec:mass_mea} in details) and the mass-to-light ratio used in ref. \cite{Reines2015}. The median offset is small ($\sim 0.01$ dex; see Figure~\ref{fig:massComparison2}) after subtracting the contribution of emission lines from broadband photometries. As the estimation error is easier to give from the SED fitting, we adopt the SED fitting with emission-line templates for stellar mass estimation.

\begin{figure}[H]
    \centering
    \includegraphics[width=0.45\textwidth]{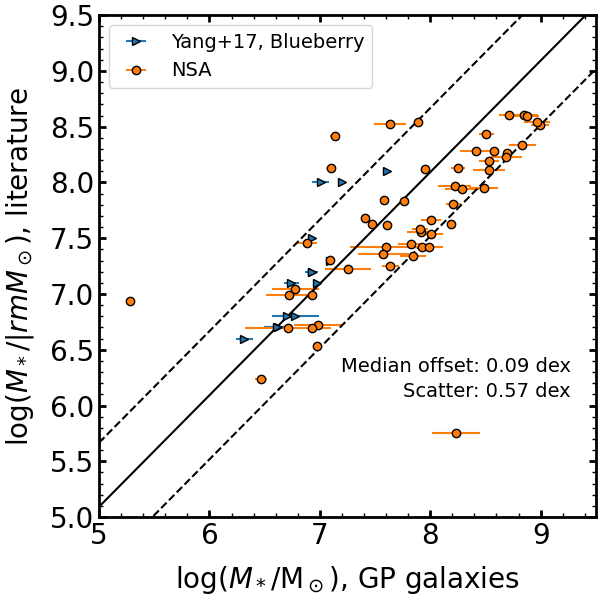}
    \caption{Comparison of stellar mass of GP parent sample derived from our SED fitting and other literature. The orange dots and blue triangles are NSA and ref.~\cite{Yang2017} samples overlapped with our GP parent galaxies, respectively. The solid and dashed lines represent the median offset and 1$\sigma$ scatter of the difference of stellar masses between our estimation and others.}
    \label{fig:massComparison}
\end{figure}

To validate the stellar mass estimation, we compare the stellar masses of our parent sample provided in the NASA-Sloan Atlas (\href{http://nsatlas.org}{NSA}) catalog based on SED fitting \cite{Blanton2007}, as well as those provided in ref.~\cite{Yang2017}. NSA estimated stellar masses using software {\tt k-correct} \cite{Blanton2007} based on ultraviolet to near-infrared data. Their SED fitting has similar SPS and IMF to ours but has different dust and emission-line models. Ref.~\cite{Yang2017} estimated the stellar masses of a sample of {\it blueberry} galaxies at $z < 0.05$. They fitted optical photometry based {\tt Starburst99} SPS model with Kroupa IMF. 68 out of 2190 GP galaxies cross-match with samples of NSA and ref.~\cite{Yang2017}. Figure~\ref{fig:massComparison} shows the comparison of stellar masses of the overlapped sample. The median offset of stellar masses between ours and the literature is 0.09 dex and the 1$\sigma$ scatter is 0.57 dex, which is similar to the uncertainty derived in the color-depended mass-to-light relation \cite{Reines2015}. As GP galaxies are blue and faint with low masses, the data quality can also introduce uncertainty.

We also check the stellar mass offset introduced by different setup parameters of SED models, such as IMF, AGN component, and ionization parameter. The stellar masses have a systematic offset of $\sim$ 0.24 dex when changing Chabrier IMF to Salpeter IMF. The median offset of stellar masses using models without AGN component is approximately equal to zero, while the offset varies with different stellar masses and has a large scatter of $\sim$ 0.33 dex. The ionization parameter is one of the determining parameters of emission-line templates. The different setups of ionization parameters introduce little offset to stellar masses, which can be negligible.

\end{appendix}

\end{multicols}
\end{document}